\documentclass[12pt]{amsart}
 \usepackage{amsmath,amssymb,amsbsy,amsfonts,amsthm,latexsym,amsopn,amstext,
 amsxtra,euscript,amscd,color}
  \newtheorem{thm}{Theorem}

  \theoremstyle{definition}
  
  \theoremstyle{remark}

  \newtheorem{question}[thm]{\bf Question}
  
  \newtheorem{prblm}[thm]{\bf Problem}

\newcommand{\cC}{{\mathcal C}}
\newcommand{\cD}{{\mathcal D}}

\newcommand{\tA}{A_B }

\renewcommand{\)}{\right)}

\newcommand{\rf}[1]{\left\lceil#1\right\rceil}

\newcommand{\lcm}{{\mathrm {lcm}}}

\begin{document}

\title[Circulant Graphs and
GCD and LCM  of Subsets]{Circulant Graphs and\\
GCD and LCM of Subsets}

\author{
  Joachim von~zur~Gathen} 
\address{B-IT, Universit{\"a}t Bonn, 
  53113 Bonn, Germany}
  \email{gathen@bit.uni-bonn.de}

\author{Igor E. Shparlinski} 
\address{Department of Pure Mathematics, University of New South Wales, 
Sydney, NSW 2052, Australia}
\email{igor.shparlinski@unsw.edu.au}

\date{\today}

\begin{abstract}
Given two sets $A$ and $B$ of integers, we consider the problem 
of finding a set $S \subseteq A$ of the smallest 
possible cardinality such the greatest common divisor of the elements 
of $S \cup B$ equals that of those of $A \cup B$. The particular cases of $B = \emptyset$ 
and $\#B = 1$ are of special interest and have some links with graph
theory.  We also consider the corresponding question for the least common multiple
of the elements. We establish NP-completeness and approximation
results for these problems by relating them to the Minimum Cover Problem. 
\end{abstract}


\maketitle

\section{Introduction}

\subsection{Description of the problem and motivation} 

For a set $A$ of integers, $\gcd(A)$ and $\lcm(A)$  denote the 
greatest common divisor ($\gcd$) and the least common multiple 
($\lcm$) of the elements of $A$, respectively. 
We consider some questions of how $\gcd$ and $\lcm$ 
behave on various subsets $S$ of the original set $A$.

We are interested in both designing algorithms to construct  
such sets $S$ with prescribed properties of  $\gcd(S)$ and $\lcm(S)$ 
and also in upper and lower bounds on what one can possibly achieve.

We consider the question of finding a subset
$S\subseteq A$ of the smallest possible cardinality with minimal
$\gcd$, namely, $\gcd(S)= \gcd(A)$, or with maximal $\lcm$, namely,
$\lcm(S) = \lcm(A)$.  We also consider a modification of this question
where we impose that a specific set $B$ of integers be contained in
$S$. This $B$ may contain elements of $A$. This question arises in the
theory of {\it circulant graphs} and is a special case of graph {\it
  editing problems\/}, see~\cite{DaMo12}, \cite{Gol13}, \cite{Math10}
and~\cite{MaSz12} for the background and further references.

To explain this connection we recall that an (undirected)
\emph{circulant graph} $G(A,m)$ on $m$ nodes, labelled $0, 1, \ldots,
m-1$, is defined by a set $A$ of integers
called {\it links\/}, where the nodes $i$ and $j$ are connected if and
only if $|i-j| \equiv a \bmod m$ for some $a \in A$.  Clearly,
$G(A,m)$ is connected if and only if $\gcd\(A\cup \{m\}\) = 1$. Thus
it is natural to ask how many links can at most be removed from $A$ so
that the new circulant graph is still connected. This leads to
the above question with $B = \{m\}$.

The above can be generalized as follows:

\begin{question} 
  \label{ques:gcd subset} Given two sets $A$ and $B$ of positive integers, find
  a subset $S \subseteq A$ of the smallest possible size with $\gcd(S
  \cup B) = \gcd(A \cup B)$.
\end{question}

Similarly, we  also ask:

\begin{question} 
  \label{ques:lcm subset} Given two sets $A$ and $B$ of positive integers, find
  a subset $S \subseteq A$ of the smallest possible size with 
$\lcm ( S \cup B) = \lcm (A \cup B)$.
\end{question}

We first formalize these questions as decision problems.

\begin{prblm} {\sl Minimum subset with minimal $\gcd$}, \textsc{MinGcd}
\begin{description}
\item[Input] Sets $A$ and $B$ of positive integers, positive integer
  $k$.
\item[Question] Does $A$ contain a subset $S$ with $\#S \leq k$ and
  $\gcd(S \cup B) = \gcd(A \cup B)$?
\end{description}
\end{prblm}

\begin{prblm} {\sl Minimum subset with maximal $\lcm$}, \textsc{MaxLcm}
\begin{description}
\item[Input] Sets $A$ and $B$ of positive integers, positive integer
  $k$.
\item[Question] Does $A$ contain a subset $S$ with $\#S \leq k$ and
  $\lcm(S \cup B) = \lcm(A \cup B)$?
  \end{description}
\end{prblm}

The input size of an instance $(A,B)$ for both  \textsc{MinGcd} and 
 \textsc{MaxLcm} is naturally defined as
$$
I(A,B) = \sum_{a\in A\cup B} \rf{\log(a+1)}
$$
where $\log z$ denotes the binary logarithm of $z\ge 1$.

For each of these (and other similar)  problems \textsc{X}, we denote as 
\textsc{OPT-X} the
corresponding optimization problem, where one has to find subsets as
described with minimal $k$.

\subsection{Main Results}
We can now formulate our main results. 

\begin{thm}
\label{thm:GCDLCM-NP}
\textsc{MinGcd} and \textsc{MaxLcm} are $\mathrm{NP}$-complete.
\end{thm}

Furthermore, a combination of the classical greedy approximation 
algorithm of~\cite[Theorem~4]{John74}
and known inapproximability results, see, for example,~\cite[Theorem~7]{AMS06},
yield the following.

\begin{thm}
\label{thm:GCDLCM-Appr}
\textsc{OPT-MinGcd} and \textsc{OPT-MaxLcm}  can be approximated in polynomial
time within a factor $O(\log I(A,B))$, but not within a factor $o(\log
I(A,B))$ if $\mathrm{P \neq NP}$.
\end{thm}

\section{Reductions Between Various Problems} 
\label{sec:Prob}

\subsection{Reduction to $B = \varnothing$}

We start by constructing from $A,B \subseteq \mathbb{Z}$ a set $\tA \subseteq
\mathbb{Z}$ so that
\begin{align*}
\text{\textsc{OPT-MinGcd}} (A,B) = \text{\textsc{OPT-MinGcd}} (\tA, \varnothing).
\end{align*}
This reduces the general case to the special situation where $B =
\varnothing$. Moreover, given a minimum solution set $S$ for one of
the two problems, one can easily find a solution for the other
one. 

For any integer $a$, we define the nonnegative integer
$$
a_{B} = \gcd(\{a\} \cup B),
$$
and apply this element-wise to any $S \subseteq \mathbb{Z}$:
$$
S_{B} = \{a_{B} \colon a \in S\}.
$$
We claim that for any $S \subseteq A$ we have
\begin{align}\label{al:SA}
\gcd(S \cup B) = \gcd(S_{B}).
\end{align}
For any $c \in \mathbb{Z}$, we have
\begin{align*}
  c \; | \gcd (S \cup B)& \Longleftrightarrow \forall a \in S \;
  \forall b  \;  \in B \quad c \;| \;a \text{ and } c \;| \;b\\
  & \Longleftrightarrow (\forall a \in S \quad c \;| \;a) \text{ and } c \;| \gcd(B)\\
  & \Longleftrightarrow \forall a \in S \quad c \;| \;a_{B}
  \Longleftrightarrow c \; | \gcd (S_{B}).
\end{align*}
In particular, we have $\gcd(A \cup B) = \gcd(A_{B})$.

Distinct $ a \in A$ may yield the same $a_{B}$.
However, if $S \subseteq A$ has minimal size with $\gcd(S \cup B) = \gcd(A
\cup B)$, then $a \mapsto a_{B}$ is injective on $S$, and $\#S =
\#S_{B}$. Thus
$$
\text{\textsc{OPT-MinGcd}} (A,B) \geq \text{\textsc{OPT-MinGcd}} (\tA, \varnothing).
$$

For the reverse direction, we take a section $\sigma$ of $a \mapsto
a_{B}$ on $A$, so that $\sigma(b) \in A$ and $(\sigma(b))_{B} = b$ for $b
\in A_{B}$.
For any $T \subseteq A_{B}$ of minimal size with $\gcd(T) = \gcd
(A_{B})$, we claim that
$$
\gcd (\sigma(T) \cup B) = \gcd (A \cup B).
$$
This follows from (\ref{al:SA}), since $(\sigma(T))_{B}= T$ and 
$$
\gcd (A \cup B) = \gcd (A_{B}) = \gcd(T) = \gcd (\sigma(T) \cup B).
$$

We have $\#\sigma(T) \leq \#T$ and $\gcd((\sigma(T))_{B}) =
\gcd(A_{B})$. The minimality of $\#T$ implies that $\#\sigma (T) =
\#T$ and thus
$$
\text{\textsc{OPT-MinGcd}} (A,B) \leq \text{\textsc{OPT-MinGcd}} (\tA, \varnothing).
$$

Overall, it follows that the minimal solution sizes for $(A, B)$ and
$A_{B}$ are equal, and that the solution sets are related by the above
correspondence. Clearly the set $A_{B}$ can be constructed in time 
polynomial in $I(A,B)$.
Thus both the decision and the optimization versions
of the general and the special cases are polynomial-time equivalent.

So from now on we assume that the input consists of one set $A$ and denote 
by $I(A) = I(A,\emptyset)$ the input size.

\subsection{Minimum Cover Problem} 

We present polynomial time reductions between \textsc{MinGcd}, \textsc{MaxLcm} and the
following problem, which is well studied in complexity theory.

\begin{prblm} {\sl Minimum cover}, \textsc{MinCover}
\begin{description}
\item[Input] List $\cC$ of subsets of a finite set $X$, positive integer
  $k$.
\item[Question] Does $\cC$ contain a cover for $X$ of size $k$ or less,
  that is, a subset $\cD \subseteq \cC$ with $\#\cD \leq k$ such that
  every element of $X$ belongs to at least one member of $\cD$?
  \end{description}
\end{prblm}

Furthermore, let $n$ be the input size, usually about $\#\cC \cdot \log m$ if $X =
\{1, \ldots, m\}$. Then \textsc{OPT-MinCover} can be approximated in
polynomial time with\-in a factor of $O(\log n)$, but no smaller
factor (unless $P=NP$), see~\cite{AMS06}. 

It is well known that \textsc{MinCover} is NP-complete, see, for
example,~\cite[Problem~SP5, Section~A.3.1]{garjoh79}.
In the next subsections, we present various reductions between
\textsc{MinCover} and our problems. The latter are trivially in NP, and their
reduction to \textsc{MinCover} transfers approximation algorithms for the latter
to approximation algorithms for our problems.
On the other hand, the reductions from \textsc{MinCover} to our problems
show that the latter cannot be approximated too well.

\subsection{Reduction from  \textsc{MaxLcm} to \textsc{MinCover}}
\label{sec:L2MC}

Let us take an instance $(A,k)$
of \textsc{MaxLcm}. We compute a \emph{coprime basis} 
($B,e$) of $A$, where $B$ consists of pairwise coprime integers $b \geq
2$ and $e  \colon~ A \times B \longrightarrow \mathbb{N}$ is such that $a =
\prod_{b \in B} b^{e(a,b)}$ for all $a \in A$. By dropping the unneeded
elements $b$ where $e(a,b)=0$ for all $a\in A$ from $B$, 
we may assume that
\begin{equation}\label{eq:BA}
\forall b \in B ~\exists ~a \in A \colon~e(a,b) \geq 1. 
\end{equation}
We recall that \cite[Section~4.8]{bacsha96} discuss coprime bases (under the
designation of \emph{gcd-free basis}) and show that one can be computed with
$O(I(A)^{2})$ bit operations, where, as before, $I(A)$ is the input size. They
use classical arithmetic. According to~\cite{bern05}, fast
arithmetic yields an algorithm using $I(A) (\log I(A))^{O(1)}$
operations. 

By the above, the size of $B$ is polynomial in that of $A$.
We note that the size of $B$ can actually be much smaller than that of $A$:
Take the first $m$ primes, all exponent vectors
$e$ in $\{1,2\}^m$, and then all $2^m$ values $ a_e = \prod_{1\leq i \leq m} p_i^{e_i}$.
Then the coprime basis $B$ consists of just these
$m$ primes and  $\text{size}(A)$ is only logarithmic in $\text{size}(B)$.
That is no worry, since we only use this reduction to derive good approximations for 
\textsc{MinCover} (which do not exist by the hardness result mentioned above) from
good approximations to our problems; hence the latter do not exist either.

For $b \in B$, we let $d(b) = \max \{e(a,b)  \colon~ a \in A\}$. Thus $d(b) \geq
1$ by~\eqref{eq:BA}, and $\lcm (A) = \prod_{b \in B} b^{d(b)}$; see 
also~\cite[Corollary~4.8.2]{bacsha96}. For $a \in A$, we set
$$
C_{a} = \{b \in B  \colon~ e(a,b)= d(b)\}.
$$
We now take a subset $E \subseteq A$ such that
$\{C_a\colon a \in A \} = \{C_a\colon a \in E \} $
and the $C_a$ in the latter set are pairwise distinct.
 Clearly this can be 
done in time polynomial in $I(A)$. It is also clear that 
$\lcm( E) = \lcm (A)$. 

Now we consider the  \textsc{MinCover}
instance with $X=B$ and $\cC=\{C_{a}  \colon~ a \in E \}$.
For $S \subseteq E$, we consider $\cD= \{C_{a}  \colon~ a \in S
\}$. Then $\#\cD = \#S$, and 
\begin{align*}
\lcm (S) = \lcm( E) = \lcm (A)  &\Longleftrightarrow \forall b \in B \quad  b^{d(b)} \mid
\lcm (S)\\
& \Longleftrightarrow \forall b \in B ~ \exists a \in S \quad b^{d(b)} \mid
a\\
& \Longleftrightarrow \forall b \in B ~ \exists a \in S \quad e(a,b) =
d(b)\\
& \Longleftrightarrow \forall b \in B = X ~ \exists a \in S \quad b \in
C_{a}\\ 
&\Longleftrightarrow \cD \text{ covers } X.
\end{align*}
Thus a solution $S$ of \textsc{MaxLcm} with $\#S \leq k$ implies one
of \textsc{MinCover} with $\#\cD \leq k$. 

Conversely,  given a cover  $\cD= \{C_{a}  \colon~ a \in S
\}$ of $X$ with $\#\cD \leq k$ we conclude that $\# S \le k$, since the sets
$C_{a}$ for $a \in E$  are pairwise distinct. 
 
Thus the size of the smallest set $S \subseteq A$ 
with $\lcm(S) = \lcm(A)$ and the size of the smallest cover 
$\cD$ of $X$ coincide. 
This concludes the reduction.

\subsection{Reduction from  \textsc{MinGcd} to \textsc{MinCover}}
\label{sec:G2MC}
We replace $d(b)$ in the previous reduction by $g(b) = \min \{e(a,b)
\colon~ a \in A\}$.
Then
$$
\gcd(A) = \prod_{b \in B}b^{g(b)}.
$$
We see that $\gcd (E) = \gcd(A)$ divides $\gcd(S)$
and in the argument for $\gcd(E) = \gcd(S)$,
 a divisibility $b^{d(b)} \mid u$ has to be replaced by $b^{g(b)+1}
\nmid u$. Otherwise the argument goes through unchanged.

\subsection{Reduction from \textsc{MinCover} to \textsc{MaxLcm}}
\label{sec:MC2L}

We are given a list $\cC$ of sets $C_{1}, \ldots, C_{l} \subseteq
X$, where $X = \{1, \ldots, m \}$, and $k \geq 1$. We may assume that  $X= \bigcup_{i \leq l} ~
C_{i}$ and $k \leq l$, otherwise the \textsc{MinCover} problem is trivial.
We let $p_{1} < p_{2} < \cdots < p_{m}$ be the first
$m$ prime numbers, $a= \prod_{j \in X} p_{j}$,
$$
a_{i} = \prod_{j \in C_{i}} p_{j},
$$
for $i \leq l$ and $A=\{a_{1}, \ldots, a_{l} \}$. Thus $a = \lcm
(A)$. We use the same value of $k$ for both problems. Since $p_{m}
\leq (1+o(1)) \, m \ln m$ as $m \to \infty$, the bit size of $(A,k)$ is in
$O(l m \log m)$. The set $A$ can be computed in
time polynomial in $lm$, using the sieve of Eratosthenes for generating
the primes.

Suppose that $I \subseteq \{1, \ldots, m \}$ is such that $\#I \leq k$
and $\lcm (S) = \lcm (A)$, where $S = \{a_{i} \colon~ i \in I \}$. Let
$$
\cD = \{C_{i}  \colon~ i \in I \}.
$$
Then $\#\cD \leq k$. Furthermore, for any $j \in X$, $p_{j}$ divides
$\lcm (A) = \lcm (S)$ and hence $a_{i}$ for some $i \in I$. It follows
that $j \in C_{i} \in \cD$. Thus $\cD$ covers $X$.

On the other hand, suppose that $I \subseteq \{1, \ldots, m\}$ is such
that $\#I \leq k$ and $\cD = \{C_{i}  \colon~ i \in I$\}
 covers $X$. Then $S = \{a_{i}  \colon~ i \in I \}$ satisfies $\#S \leq
 k$ and $\lcm (S) = a = \lcm (A)$.

\subsection{Reduction from \textsc{MinCover} to \textsc{MinGcd}}
\label{sec:MC2G}
For an analogous reduction to \textsc{MinGcd}, we replace
$a_{i}$ by $a/p_{i}$ in the above.

\section{Proofs of Main Results}

We start with the upper bounds claimed in Theorems
\ref{thm:GCDLCM-NP} and \ref{thm:GCDLCM-Appr}.
The fact that \textsc{MaxLcm} and \textsc{MinGcd} are NP-complete is trivial.
Furthermore, the reductions of Sections~\ref{sec:L2MC} and~\ref{sec:G2MC}
show that the know approximation algorithms for \textsc{MinCover}
also yield ones for our problems.

Furthermore, our claimed lower bounds (NP-hardness and inapproximability)
follow from the reductions in
Sections~\ref{sec:MC2L} and~\ref{sec:MC2G}
from our problems to
\textsc{MinCover}, together with the NP-hardness and inapproximability
of \textsc{MinCover}, as cited above.

\section*{Acknowledgments}

The authors are grateful to Fedor Fomin for some references on 
the approximability and inapproximability of the set-cover problem. 

The first author's work was supported by the B-IT Foundation and the
Land Nordrhein-Westfalen. The second author's work was supported by the 
Australian Research Council grants DP110100628 and DP140100118.

\end{document}